\documentclass[sigconf]{acmart}

\usepackage{subcaption}  
\usepackage{array}
\newcolumntype{L}[1]{>{\raggedright\let\newline\\\arraybackslash\hspace{0pt}}m{#1}}
\newcolumntype{C}[1]{>{\centering\let\newline\\\arraybackslash\hspace{0pt}}m{#1}}
\newcolumntype{R}[1]{>{\raggedleft\let\newline\\\arraybackslash\hspace{0pt}}m{#1}}

\usepackage{xfrac}
\usepackage{nicefrac}
\usepackage{makecell}

\usepackage{amsopn}

\usepackage{xspace,fancyvrb}
\usepackage{multirow}

\usepackage{enumitem}
\usepackage{url}

\usepackage{fp}
\usepackage{siunitx}

\usepackage{listings}
\definecolor{codegreen}{rgb}{0,0.6,0}
\definecolor{codegray}{rgb}{0.5,0.5,0.5}
\definecolor{codepurple}{rgb}{0.58,0,0.82}
\definecolor{backcolour}{rgb}{0.95,0.95,0.92}

\lstdefinestyle{mystyle}{
    commentstyle=\color{codegreen},
    keywordstyle=\color{magenta},
    numberstyle=\tiny\color{codegray},
    stringstyle=\color{codepurple},
    basicstyle=\ttfamily\footnotesize,
    breakatwhitespace=false,         
    breaklines=true,                 
    captionpos=b,                    
    keepspaces=true,                 
    numbers=left,                    
    numbersep=5pt,                  
    showspaces=false,                
    showstringspaces=false,
    showtabs=false,                  
    tabsize=2
}

\usepackage{framed}
\usepackage{xcolor}
\definecolor{shadecolor}{gray}{0.95}
\newenvironment{findingbox}{
   \setlength{\fboxsep}{6pt}
   \setlength{\fboxrule}{0.5pt}
   
   \MakeFramed {\advance\hsize-\width \FrameRestore}
   \noindent\textbf{Finding:}
}{
   \endMakeFramed
}

\ccsdesc[500]{Security and privacy~Software security engineering}

\keywords{Software Security, Automated Vulnerability Repair, Large Language Models, Test-based Evaluation}

\copyrightyear{2026}
\acmYear{2026}
\setcopyright{cc}
\setcctype{by-nc-nd}
\acmConference[ICSE '26]{2026 IEEE/ACM 48th International Conference on Software Engineering}{April 12--18, 2026}{Rio de Janeiro, Brazil}
\acmBooktitle{2026 IEEE/ACM 48th International Conference on Software Engineering (ICSE '26), April 12--18, 2026, Rio de Janeiro, Brazil}
\acmPrice{}
\acmDOI{10.1145/3744916.3773167}
\acmISBN{979-8-4007-2025-3/2026/04}

\begin{document}

\title{Rethinking the Capability of Fine-Tuned Language Models for Automated Vulnerability Repair}

\settopmatter{authorsperrow=4}

\author{Woorim Han}
\affiliation{%
  \institution{Seoul National University}
  \city{Seoul}
  \country{South Korea}
}
\email{rimwoohan@gmail.com}

\author{Yeongjun Kwak}
\affiliation{%
  \institution{UNIST}
  \city{Ulsan}
  \country{South Korea}}
\email{kyj05137@unist.ac.kr}

\author{Miseon Yu}
\affiliation{%
  \institution{Seoul National University}
  \city{Seoul}
  \country{South Korea}}
\email{miseon543@gmail.com}

\author{Kyeongmin Kim}
\affiliation{%
 \institution{UNIST}
 \city{Ulsan}
 \country{South Korea}}
\email{kim1254@unist.ac.kr}

\author{Younghan Lee}
\authornote{Corresponding author.}
\affiliation{%
  \institution{Sungshin Women's University}
  \city{Seoul}
  \country{South Korea}}
\email{yhlee@sungshin.ac.kr}

\author{Hyungon Moon}
\affiliation{%
  \institution{UNIST}
  \city{Ulsan}
  \country{South Korea}}
\email{hyungon@unist.ac.kr}

\author{Yunheung Paek}
\authornotemark[1]
\affiliation{%
  \institution{Seoul National University}
  \city{Seoul}
  \country{South Korea}}
\email{ypaek@snu.ac.kr}

\renewcommand{\shortauthors}{Han et al.}

\renewcommand{\subsectionautorefname}{\S}
\renewcommand{\sectionautorefname}{\S}
\newcommand{\PP}[1]{\noindent\textbf{#1.}\xspace}

\begin{abstract}
Learning-based automated vulnerability repair (AVR) techniques that utilize fine-tuned language models have shown promise in generating vulnerability patches. 
However, questions remain about their ability to repair unseen vulnerabilities.
Our empirical study reveals that state-of-the-art models often overfit to the training set and are evaluated using training, validation, and test sets that are not mutually exclusive.
Furthermore, relying on match-based metrics that compare generated patches to reference fixes at the token level has some limitations, failing to account for the possibility of various valid ways to patch the vulnerability.
In this paper, we examine the capabilities of state-of-the-art fine-tuned AVR models and the adequacy of match-based evaluation metrics in three ways.
First, we apply semantic-preserving transformations to test sets in order to determine whether models truly learn robust vulnerability-repair patterns or simply rely on spurious features.
Second, we re-split the training, validation, and test sets to be mutually exclusive and evaluate the models on the revised test set to assess their generalization capabilities.
Third, we introduce L-AVRBench, a test-based benchmark tailored for learning-based AVR, to overcome the limitations of match-based metrics and examine the AVR models' true repair capabilities.
\end{abstract}

\maketitle

\section{Introduction}
\begin{figure}[tb]
    \centering
        \begin{subfigure}{1\columnwidth}
        \centering
        \includegraphics[width=\linewidth]{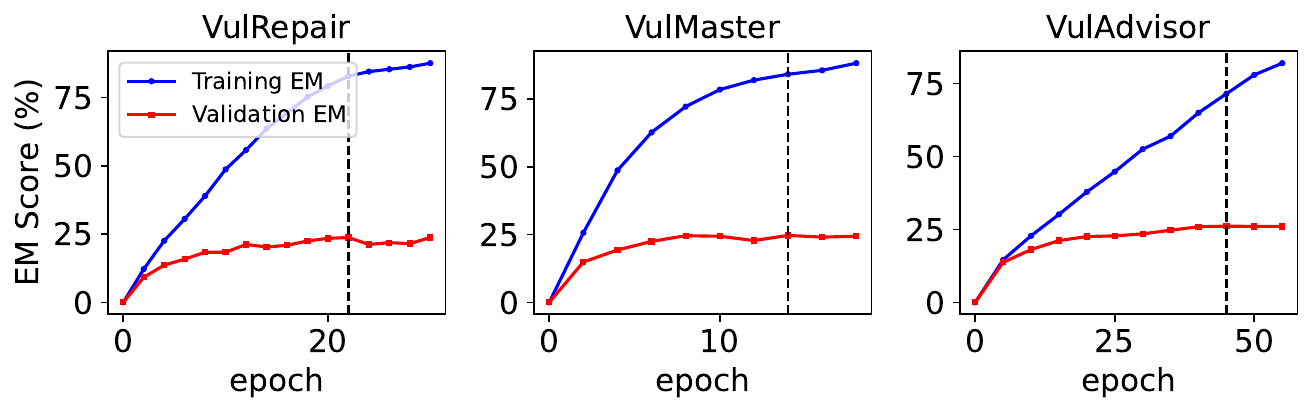}
        \caption{Training and validation EM. The widening gap between training and validation EM suggests overfitting.}
        \label{fig: mot-em}
    \end{subfigure}

    \begin{subfigure}{1\columnwidth}
        \centering
        \includegraphics[width=\linewidth]{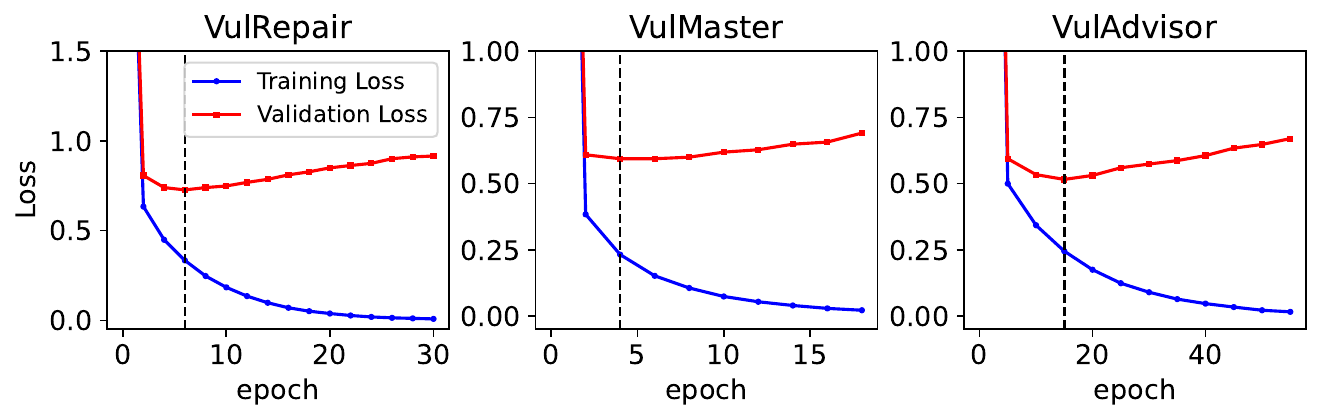}
        \caption{Training and validation loss curves. The increasing validation loss despite rising validation EM suggests potential data overlap.}
        \label{fig: mot-loss}
    \end{subfigure}
    \Description{Learning curves for SOTA Fine-tuned language models suggesting overfitting and possible overlapping data problem.}
    \caption{Learning curves for SOTA Fine-tuned language models suggesting overfitting and possible overlapping data problem. The black vertical lines indicate the best validation EM performance (a) and lowest validation loss (b).}
    \label{fig: mot}
\end{figure}

With rapid advancements in the software landscape, the prevalence of security vulnerabilities is rising, posing significant risks to users and organizations~\cite{forbes2023zeroday,shin2010evaluating}. In 2023, over 28,000 Common Vulnerabilities and Exposures (CVEs) were reported, which was a 15\% increase from 2022~\cite{gamblin2024cve}, and by August 2024, over 52,000 new CVEs had already been identified, setting a new record~\cite{statista2024cve}. 
Despite this escalation, patching vulnerabilities remains labor-intensive and requires specialized expertise, leaving many vulnerabilities unpatched or mitigated only after significant delays~\cite{huang2016talos,li2017large,ji2018coming}.

Automated Vulnerability Repair (AVR) has emerged as a promising solution to accelerate the patching process and mitigate security risks by automatically generating fixes.
Early AVR approaches~\cite{gao2019crash,gao2021beyond,hong2020saver,huang2019using,lee2018memfix, ma2016cdrep, van2018static}, such as program analysis and search-based methods, have shown effectiveness in generating patches for specific kinds of vulnerabilities.
The rise of large language models (LLMs) has also arrived in the AVR domain, leading to the development of many learning-based AVR models~\cite{liu2024t, vulrepair,vulmaster,zhang2024vuladvisor,pearce2023examining, liu2024exploring}.
These approaches often fine-tune pre-trained language models on vulnerability datasets~\cite{liu2024t, vulmaster, vulrepair} containing various types of vulnerabilities, enabling them to handle a wider range of security flaws without being restricted to specific vulnerability types.

Recent studies show that fine-tuned AVR models can generate effective patches, but our in-depth analysis, shown in~\autoref{fig: mot}, suggests that existing AVR models may have limited generalization capabilities.
The figure illustrates the learning curves of the state-of-the-art (SOTA) fine-tuned AVR models, and the curves show that the model chosen (black dotted line) has the generalization gap~\cite{jiang2018predicting, johnson2023inconsistency}. 
Specifically, all fine-tuned AVR models attain over 70\% training Exact Match (EM), yet their validation EM plateaus at a much lower level.
This discrepancy suggests that the model overfit to the training set, relying on spurious features, such as identifier names and function parameters, rather than truly understanding the vulnerability.
We also observe that the validation EM continues to rise even as the validation loss begins to increase before epoch 5 (\autoref{fig: mot-loss}).
Such a phenomenon has been linked to~\textit{overlapping data} between training and validation sets, which are identical or highly similar examples appearing in both partitions~\cite{elangovan2021memorization, Gorman2019, Recht2019}.
The overlapping data causes the model to inadvertently encounter validation instances during training, inflating its validation performance without truly learning new vulnerability patterns and thus increasing validation EM despite a rising validation loss.
These findings highlight the need for a deeper examination of fine-tuned AVR models, as accurately assessing their true ability to repair is crucial for understanding their real-world effectiveness and guiding future research in that direction.

However, the current evaluation approach for fine-tuned AVR models has a limitation in that their performance is mainly judged by comparing the generated patches to a single reference solution.
This evaluation is typically conducted using EM (Exact Match) or fuzzy matching (BLEU~\cite{papineni2002bleu}, CodeBLEU~\cite{ren2020codebleu}).
However, studies across various sequence generation tasks (e.g., NLP, machine translation, question answering, and code generation) have demonstrated that these match-based metrics may not accurately reflect the true performance of the model~\cite{evtikhiev2023out, chen2021evaluating, lukasik2024regressionawareinferencellms, bulian2022tomayto, schaeffer2023emergent}.
EM penalizes the generated answers that are not exactly the same but semantically equivalent to the reference~\cite{schaeffer2023emergent, lukasik2024regressionawareinferencellms,bulian2022tomayto}.
Similarly, match-based metrics fail to account for the many functionally equivalent solutions possible in code-related tasks~\cite{opitz2023codebertscore, chen2021evaluating, dong2023codescore, haque2022fixeval}.
Recognizing these limitations, recent studies in code generation and program repair have introduced test-based evaluation methods that assess a model's effectiveness by verifying whether its generated code passes functionality test cases~\cite{wang2023rap, just2014defects4j, austin2021program, hendrycks2021measuring}.
Despite these advancements in related fields, the limitation of match-based evaluation remains largely unexamined in the context of AVR.
Given that AVR requires not only syntactic correctness but also functional correctness and security guarantees, there remains an open question about how well these match-based metrics capture the true quality of generated patches and the model's actual repair capabilities when evaluated with a test-based method.

In this paper, we seek to clarify these questions by systematically examining both the performance of existing fine-tuned AVR models and the adequacy of current evaluation metrics. To thoroughly evaluate the capabilities of fine-tuned AVR models and investigate whether they rely on spurious features rather than genuine vulnerability-repair patterns, we apply semantic-preserving transformations to the test set. This is motivated by prior studies showing that learning-based models often struggle when tested on semantically transformed code, suggesting that they may overfit to surface-level patterns rather than truly understanding the code~\cite{applis2021assessing, bielik2020adversarial,rabin2021generalizability, risse2024uncovering}.
In addition, we study the impact of overlapping data across training, validation, and test sets, evaluating AVR models with a chronological data split that resolves the overlapping problem and ensures a more accurate assessment of their repair capabilities.

We also introduce L-AVRBench, a test-based benchmark specifically tailored for learning-based AVR models to assess whether current metrics effectively evaluate the quality of vulnerability patches and to evaluate the true repair capabilities of fine-tuned AVR models beyond match-based metrics.
Our benchmark refines the existing AVR benchmark ExtractFix~\cite{gao2021beyond, nusapr2022EFbench}, which predates the prevalence of learning-based AVR models and is not tailored for them, to ensure that the curated data remains suitable for evaluating learning-based AVR models.
Since this adaptation reduces the number of usable samples, we augment the benchmark with additional cases from Magma~\cite{hazimeh2020magma}, a ground-truth fuzzing benchmark, and hand-mined CVE-tagged vulnerabilities from public bug reports~\cite{nvdsite, ossfuzz}. 
These additions increase both dataset size and vulnerability diversity, ensuring that L-AVRBench captures a wide and realistic spectrum of security flaws.
Our study newly raises and answers the following four research questions.

\textbf{RQ1: Do the fine-tuned AVR models effectively learn vulnerability patterns?}
(\autoref{sec:3.1})
Our analysis shows that fine-tuned AVR models tend to memorize spurious features of vulnerable functions.
When test data is augmented using semantic-preserving transformations, performance drops across models, with EM decreasing by 4.0\%, BLEU by 4.5\%, and CodeBLEU by 7.0\% on average.
   
\textbf{RQ2: Is the current experimental setting suitable for accurately evaluating model performance?}
(\autoref{sec:3.3})
Our analysis shows that random data splitting in AVR inflates performance due to train–test overlap. Under chronological splitting that eliminates overlap, VulRepair’s EM drops from 18.3\% to 4.96\%, and VulMaster’s from 20.47\% to 5.15\%.

\textbf{RQ3: How well do match-based metrics (i.e., EM, BLEU, CodeBLEU) correlate with key aspects of vulnerability repair, such as functional correctness and security?}
(\autoref{ss:rq3})
We find that exact-match evaluation, while guaranteeing correctness, fails to acknowledge valid alternative fixes and becomes unreliable when reference patches rely on external variables or function calls.
Furthermore, BLEU and CodeBLEU exhibit a low correlation with the security and functional correctness of the generated patches.

\textbf{RQ4: What is the capability of fine-tuned AVR models when evaluated with a test-based benchmark (Our L-AVR Bench)?}
(\autoref{sec5})
When assessed with L-AVRBench, VulRepair, VulMaster, and VulAdvisor successfully repair 4, 15, and 5 out of 70 vulnerability IDs, respectively.
Notably, only one instance achieved an EM score of 1, highlighting the rarity of exact-match patches.

In summary, we make the following contributions:
\begin{itemize}[leftmargin=*,labelsep=0.5em]
\item We empirically demonstrate that SOTA fine-tuned AVR models overfit to spurious features that are unrelated to the underlying vulnerability and show that the current evaluation setting inflates the true repair capability of these models.

\item We are the first to go beyond match-based metrics in evaluating fine-tuned AVR models for C/C++ vulnerabilities, demonstrating their limitations in capturing key aspects of repair quality, such as functional correctness and security guarantees.

\item We introduce L-AVRBench, a test-based benchmark for learning-based AVR, consisting of 70 vulnerable C/C++ functions with executable test suites for evaluating patch correctness and security.

\end{itemize}

\begin{table*}[t]
    \centering
    \renewcommand{\arraystretch}{1.2} 
    \footnotesize
    \setlength{\tabcolsep}{5pt}        
    \caption{Overview of the SOTA AVR models, including their descriptions, reproduced test results, and previously reported test results.\protect\footnotemark \ B and CB denote BLEU and CodeBLEU, respectively. The Used Metrics column shows the evaluation metrics used in the original papers.}
    \label{tab:tab1}
    \centering
\renewcommand{\arraystretch}{1.1} 
\resizebox{\textwidth}{!}{
\begin{tabular}{|c|ccccc|m{0.5cm}<{\centering}m{0.5cm}<{\centering}m{0.5cm}<{\centering}|m{0.5cm}<{\centering}m{0.5cm}<{\centering}m{0.5cm}<{\centering}|}
    \hline
    \multirow{2}{*}{Evaluated Models} & \multicolumn{5}{c|}{Model Description} & \multicolumn{3}{c|}{Reported Results} & \multicolumn{3}{c|}{Reproduced Results}  \\
    & Dataset  & Model & Input & Output & Used Metrics & EM & B & CB & EM & B & CB\\ 
    \hline\hline
    
    \centering{VulRepair~\cite{vulrepair}} & BigVul+CVEFixes & CodeT5 & Tokenized function & Token context diff & EM  & 16.80 & 24.20 & 35.30 & 18.30 & 56.60 & 55.57\\ 
    
    \centering{VulMaster~\cite{vulmaster}} & BigVul+CVEFixes & CodeT5 (FiD) & Tokenized function & Token context diff & EM, B, CB & 20.00 & 29.30 & 40.90 & 20.47 & 59.81 & 58.25  \\  
   
    \centering{VulAdvisor~\cite{zhang2024vuladvisor}} & VulAdvisor & CodeT5 & Function with newline & Diff & EM, B & 22.80 & 61.10 & - & 23.33 & 60.00 & 57.50  \\  
    \hline
\end{tabular}
}

\end{table*}
    \footnotetext{The reported test results for the VulRepair model are taken from the VulMaster paper~\cite{vulmaster}.}
    
\section{Backgrounds and Related Work}
\subsection{Fine-tuned AVR Models}
Research on AVR using pre-trained Language Models treats the task as a sequence-to-sequence problem, where a vulnerable code snippet $X_i$ is transformed into its repaired version $Y_i$.
Specifically, given a vulnerable function $X_i$, the model generates the corresponding code patch $Y_i$.

\PP{VulRepair} 
VulRepair~\cite{vulrepair} is the first to fine-tune a pre-trained language model specifically for the task of AVR.
By utilizing a fine-tuned version of the CodeT5 model~\cite{codet5}, VulRepair aims to automatically generate patches for vulnerabilities in C/C++ systems.
VulRepair demonstrates the potential of using pre-trained language models in AVR by achieving competitive performance on real-world vulnerability datasets~\cite{cvefixes, bigvul}.
Its design enables the model to understand code context and generate fixes by taking vulnerable code as input and predicting the appropriate patch.

\PP{VulMaster} 
VulMaster~\cite{vulmaster} is an AVR model designed to address the limitations of existing models~\cite{vrepair,vulrepair}, particularly the input length constraints in transformer-based architectures.
It leverages the Fusion-in-Decoder (FiD) framework~\cite{izacard2020leveraging} to effectively process long and complex vulnerable code, surpassing typical model limitations.
To enhance the ability to repair, VulMaster integrates Abstract Syntax Trees (ASTs) to preserve the semantic structure of code, incorporates CWE (Common Weakness Enumeration)~\cite{CWE2023} knowledge, and utilizes vulnerability-fix examples for both the same and related CWE-IDs.
By combining these elements with the vulnerable input, VulMaster generates more accurate repairs.
This design enables VulMaster to achieve higher EM, BLEU, and CodeBLEU scores compared to previous approaches, demonstrating improved effectiveness in vulnerability repair.

\PP{VulAdvisor} VulAdvisor~\cite{zhang2024vuladvisor} is a developer-centric AVR system that provides natural language repair suggestions instead of directly generating patches.
Unlike the other AVR models that focus solely on automated patch generation, VulAdvisor guides developers through step-by-step remediation, enhancing both interpretability and trust in vulnerability fixes. 
Although its core contribution lies in an LLM-based vulnerability repair suggestion model, we include VulAdvisor in our study because it demonstrates SOTA repair effectiveness in EM and BLEU when natural language suggestions are integrated into the vulnerable code.

\subsection{Reproducing the AVR Models}
As shown in~\autoref{tab:tab1}, we reproduce two SOTA fine-tuned AVR models, VulMaster~\cite{vulmaster} and VulAdvisor~\cite{zhang2024vuladvisor}, and their baseline VulRepair ~\cite{vulrepair}.
To ensure the preciseness and fairness of the analysis, we carefully reproduce the models by training and validating them using the same datasets and data splits as reported in their original paper.
For VulRepair and VulMaster, we replicated the training process as described in the VulMaster paper, utilizing a merged and deduplicated dataset from CVEFixes~\cite{cvefixes} and Big-Vul~\cite{bigvul}, where 60\% of duplicate pairs were removed to prevent inflated performance results.
Given that VulMaster undergoes a pretraining phase with a general bug-fix corpus before fine-tuning on vulnerability-fix pairs, we also reproduced VulRepair following the same pretraining strategy to maintain consistency.
Additionally, we reproduced VulAdvisor, ensuring that its dataset aligns with the large-scale vulnerability-fix pairs collected by its authors from real-world C/C++ projects.
However, instead of using VulAdvisor's fine-tuned suggestion model for suggestion-based repair, we opted to use Ground Truth suggestions, as they have been reported to yield higher EM and BLEU.
For all three models, we stopped training when the validation performance (i.e., EM) converged.

Furthermore, we carefully followed the input preprocessing and output representation of each model. While VulRepair and VulMaster preprocess inputs into tokenized functions and use token context diff~\cite{vrepair} as the output format, VulAdvisor employs a different approach. Specifically, VulAdvisor retains vulnerable functions with newlines (e.g., line breaks such as \textbackslash n), without tokenizing them. Moreover, rather than using token context diff as the model’s output, VulAdvisor instead utilizes Git diff to represent the required changes. By maintaining these distinct preprocessing and output strategies for each model, we ensured that our reproduction adhered to the original methodologies, enabling a fair and direct comparison of performance across models.

Our reproduced EM results deviate from the original papers by no more than 2\%.
We note that when computing BLEU and CodeBLEU, our analysis of their implementation revealed that VulMaster computes these metrics by treating all generated outputs as a single corpus.
This approach can introduce differences in precision due to global aggregation effects and length normalization at the corpus level.
In contrast, we adopted VulAdvisor's method of calculating these scores on a per-sample basis, which resulted in an increase of approximately 20 percentage points.
We believe this per-sample approach is more appropriate, as it evaluates each vulnerable function independently, preventing potential distortions that arise when merging outputs with different characteristics into a single corpus.

\subsection{Existing Evaluation Metrics for Fine-tuned AVR Models}
Accurately assessing the performance of fine-tuned AVR models is essential for advancing their capabilities. 
Ideally, the evaluation of generated patches needs a comprehensive set of test cases to assess their correctness, as there can be multiple appropriate fixes for a single vulnerability.
However, obtaining real-world vulnerabilities along with comprehensive test cases is highly challenging. For this reason, previous studies often compare the generated patches against developer-provided patches as a proxy for correctness.
To facilitate this comparison, commonly used evaluation metrics are adapted from related domains, such as machine translation and code generation, to measure the quality and relevance of the generated patches.
Below, we discuss some metrics widely used to evaluate AVR models.

\PP{Exact Match}
Exact Match (EM) is a simple yet widely used metric, originally employed in natural language tasks such as question answering~\cite{rajpurkar2016squad}. It measures whether a generated output exactly matches the ground truth. Formally, \( EM = 1 \) if the generated output is identical to the ground truth at the character level; otherwise, \( EM = 0 \). The overall EM score is computed as the fraction of samples with  \( EM = 1 \) across the evaluation dataset.
While EM is straightforward and easy to compute, its applicability to code-related tasks is limited because it strictly penalizes functionally equivalent code that differs syntactically from the ground truth.

\PP{BLEU}
Papineni et al.~\cite{papineni2002bleu} introduced the Bilingual Evaluation Understudy (BLEU) metric to evaluate machine-generated translations~\cite{koehn2003statistical} by measuring n-gram overlap between a Candidate and one or more Reference(s), with a brevity penalty for overly short outputs. Although originally designed for natural language, BLEU is widely used in code-related tasks due to its simplicity and scalability. However, it is limited for security evaluation because it ignores the syntactic structure and semantics of code.

\PP{CodeBLEU}
Recognizing the limitations of BLEU for code evaluation, Ren et al.~\cite{ren2020codebleu} proposed CodeBLEU, a metric tailored for machine-generated code. CodeBLEU extends BLEU to account for three major characteristics of source code: (1) a finite set of keywords leading to repetitive patterns, (2) hierarchical structure represented by Abstract Syntax Trees (ASTs), and (3) stricter semantic requirements with little tolerance for ambiguity. Accordingly, CodeBLEU incorporates both AST-based structural similarity and data-flow information to better assess generated patches. Although widely adopted for its ability to capture syntactic and semantic properties of code~\cite{ahmad2021unified}, CodeBLEU remains a match-based metric and does not directly measure functional correctness.
\section{The Capabilities of Fine-tuned AVR Models} 

\PP{Overview}
This section presents two key observations, supported by empirical evidence, about the training and evaluation of recently introduced fine-tuned AVR models.
In particular, we observe a significant generalization gap, as these models achieve high performance on the training set but struggle on validation sets, suggesting potential overfitting.
Further analysis reveals that there is overlapping data between the training, validation, and test sets, and this overlap leads to inflated performance estimates.
We validate the first observation by assessing if the models truly learn the vulnerability patterns or rely on superficial syntactic features (\textbf{RQ1}).
For the second observation, we re-split the data to remove overlap to train and evaluate the models of interest under this non-overlapping setting (\textbf{RQ2}).

\subsection{Overfitting Problem (RQ1)}
\label{sec:3.1}
\begin{table}[t]
    \centering
    \renewcommand{\arraystretch}{1.2} 
    \setlength{\tabcolsep}{5pt}        
    
    \footnotesize
    \caption{EM, BLEU, and CodeBLEU score for VulRepair(VR), VulMaster(VM), and VulAdvisor(VA) after applying semantic-preserving transformations to the test set. Each cell shows the new score, followed by the score drop (in parentheses) relative to the original (No Transformation).}
    \label{tab:trans}
\begin{tabular}{|c|C{1.23cm}|C{1.63cm}|C{1.63cm}|c|}
    \hline
    \textbf{TF}  & \textbf{Metric} & \textbf{VR} & \textbf{VM} & \textbf{VA} \\
    \hline \hline

    \multirow{3}{*}{\makecell{No \\ TF}} 
        & EM        & 18.30  & 20.47  & 23.33 \\
            & BLEU      & 56.60  & 59.81  & 60.00 \\
            & CodeBLEU  & 55.57  & 58.25  & 57.50 \\
    \hline

    \multirow{3}{*}{\( t_1 \)} 
    & EM        & 11.97 (-6.33)  & 12.47 (-8.00)  & 15.24 (-8.09) \\
                   & BLEU      & 51.87 (-4.74)  & 53.92 (-5.89)  & 50.83 (-9.17) \\
                   & CodeBLEU  & 51.43 (-4.14)  & 54.09 (-4.19)  & 49.52 (-7.98) \\
    \hline

    \multirow{3}{*}{\( t_2 \)} 
        & EM        & 14.70 (-3.60)  & 16.50 (-3.97)  & 17.66 (-5.67) \\
                   & BLEU      & 53.14 (-3.46)  & 55.81 (-4.00)  & 51.98 (-8.02) \\
                   & CodeBLEU  & 53.39 (-2.18)  & 55.89 (-2.36)  & 50.93 (-6.57) \\
    \hline

    \multirow{3}{*}{\( t_3 \)} 
        & EM        & 13.83 (-4.47)  & 15.82 (-4.65)  & 18.27 (-5.06) \\
                   & BLEU      & 53.03 (-3.57)  & 55.99 (-3.82)  & 52.21 (-7.79) \\
                   & CodeBLEU  & 52.76 (-2.81)  & 55.49 (-2.76)  & 51.31 (-6.19) \\
    \hline

    \multirow{3}{*}{\( t_4 \)} 
        & EM        & 13.46 (-4.84)  & 15.14 (-5.33)  & 15.68 (-7.65) \\
                   & BLEU      & 53.19 (-3.41)  & 55.25 (-4.56)  & 49.70 (-10.30) \\
                   & CodeBLEU  & 52.89 (-2.76)  & 55.24 (-3.01)  & 49.04 (-8.46) \\
    \hline

    \multirow{3}{*}{\( t_5 \)} 
        & EM        & 14.70 (-3.60)  & 16.19 (-4.28)  & 18.22 (-5.11) \\
                   & BLEU      & 53.55 (-3.05)  & 55.94 (-3.87)  & 52.08 (-7.92) \\
                   & CodeBLEU  & 53.10 (-2.47)  & 55.49 (-2.76)  & 51.08 (-6.42) \\
    \hline

    \multirow{3}{*}{\( t_6 \)} 
        & EM        & 12.34 (-5.96)  & 14.08 (-6.39)  & 17.35 (-5.97) \\
                   & BLEU      & 51.66 (-4.94)  & 54.48 (-5.33)  & 50.96 (-9.04) \\
                   & CodeBLEU  & 51.67 (-3.90)  & 54.17 (-4.08)  & 50.28 (-7.22) \\
    \hline

    \multirow{3}{*}{\( t_8 \)} 
        & EM        & 13.15 (-5.15)  & 14.95 (-5.52)  & 16.00 (-7.33) \\
                   & BLEU      & 52.97 (-3.63)  & 54.46 (-5.35)  & 49.76 (-10.24) \\
                   & CodeBLEU  & 52.55 (-3.02)  & 55.21 (-3.04)  & 49.21 (-8.29) \\
    \hline

    \multirow{3}{*}{\( t_{10} \)} 
        & EM        & 14.70 (-3.60)  & 16.19 (-4.28)  & 18.22 (-5.11) \\
                   & BLEU      & 53.35 (-3.05)  & 55.94 (-3.87)  & 52.08 (-7.92) \\
                   & CodeBLEU  & 53.09 (-2.48)  & 55.49 (-2.76)  & 51.12 (-6.38) \\
    \hline
    
    \multirow{3}{*}{\( t_{11} \)} 
        & EM        & 13.77 (-4.53)  & 14.52 (-5.95) & 17.36 (-5.97) \\
        & BLEU      & 53.12 (-3.48)  & 54.22 (-5.59)  & 51.60 (-8.40)  \\
        & CodeBLEU  & 52.72 (-2.85)  & 55.67 (-2.58)  & 51.44 (-6.06) \\
    \hline

    \multicolumn{5}{l}{No TF: Reproduced SOTA Models (No Transformation Applied)} \\
    \multicolumn{5}{l}{$t_1$: Identifier Renaming (Rename all function parameters to a random token)} \\
    \multicolumn{5}{l}{$t_2$: Stmt Reordering (Reorder all function parameters)} \\
    \multicolumn{5}{l}{$t_3$: Identifier Renaming (Rename the function)} \\
    \multicolumn{5}{l}{$t_4$: Stmt Insertion (Insert unexecuted code)} \\
    \multicolumn{5}{l}{$t_5$: Stmt Insertion (Insert comment)} \\
    \multicolumn{5}{l}{$t_6$: Stmt Reordering (Move the code of the function into a separate function)} \\
    \multicolumn{5}{l}{$t_8$: Stmt Insertion (Define additional void function and call it from the function)} \\
    \multicolumn{5}{l}{$t_{10}$: Stmt Insertion (Add code from training set as comment)} \\
    \multicolumn{5}{l}{$t_{11}$: \makecell[l]{Random Transformation (Apply one random transformation from $t_1$–$t_{10}$ \\ per function)}} \\
\end{tabular}

\end{table}

To evaluate whether learning-based AVR models learn to generate patches based on semantic understanding rather than spurious features, we leverage a suite of semantic-preserving transformations.
These transformations alter the syntactic structure of code, such as identifier names, formatting, or statement order, without changing its semantics or observable behavior.
Our rationale for adopting these transformations is based on a key premise: vulnerability repair should depend on understanding the semantics of the code, not on memorized syntactic forms.
In other words, a robust AVR model should produce consistent repairs for code variants that are semantically equivalent.

We adopt the transformation set proposed in prior work on vulnerability detection models~\cite{risse2024uncovering}, which was carefully curated to preserve code functionality and naturalness~\cite{naturalness}. We observed that these properties were similarly preserved in our AVR dataset, reinforcing these transformations as a principled and controlled way to probe overfitting in AVR models.

\PP{Experimental Details}
We employ 9 transformations, each designed to modify the syntactic structure of the code while preserving its semantics.
This allows us to determine whether observed performance drops are due to a model's reliance on superficial syntactic patterns rather than its ability to understand and repair vulnerabilities at a deeper semantic level.
\textit{Identifier renaming} \((t_1, t_3)\) evaluates the model’s generalization capability by replacing variable and function names with semantically meaningless alternatives. If performance degrades, it may suggest that the model is overfitting to seen identifiers rather than learning underlying logic.
\textit{Statement reordering} \((t_2, t_6)\) evaluates robustness to stylistic variation. This could assess whether the model can remain accurate when code structure changes in ways that do not affect functionality. 
\textit{Statement insertion} \((t_4, t_5, t_8, t_{10})\) test sensitivity to non-functional content. Inserting non-functional or irrelevant code helps reveal whether the model can ignore misleading or noisy information. It checks whether the model maintains focus on semantically important code.
Finally, \textit{random transformation} \((t_{11})\) simulates real-world variability and assesses aggregate robustness.
Although 11 transformations were originally considered for vulnerability detection models~\cite{risse2024uncovering}, we excluded comment removal, as the training data for VulRepair and VulMaster does not contain any comments. We also excluded white space insertion because duplicate spaces are removed during the AVR models' preprocessing stage~\cite{vulrepair,vulmaster,zhang2024vuladvisor}.

\PP{Results}
As summarized in~\autoref{tab:trans}, all three learning-based AVR models (VulRepair, VulMaster, and VulAdvisor) experience a significant drop in EM, BLEU, and CodeBLEU scores when semantic-preserving transformations are applied. 
This decline indicates that the models are overly sensitive to superficial aspects of the source code. Instead of consistently repairing the underlying vulnerability, they frequently fail when minor stylistic or structural modifications are introduced, even though the semantic meaning remains unchanged.


\begin{findingbox}
The substantial performance drop observed after applying semantic-preserving transformations indicates that fine-tuned AVR models are highly sensitive to spurious features. This suggests that existing AVR models rely heavily on memorized token sequences rather than learning robust vulnerability repair patterns.
\end{findingbox}

\begin{figure}[t]
    \centering
    \begin{subfigure}[b]{0.45\textwidth}
        \centering
        \includegraphics[width=\textwidth]{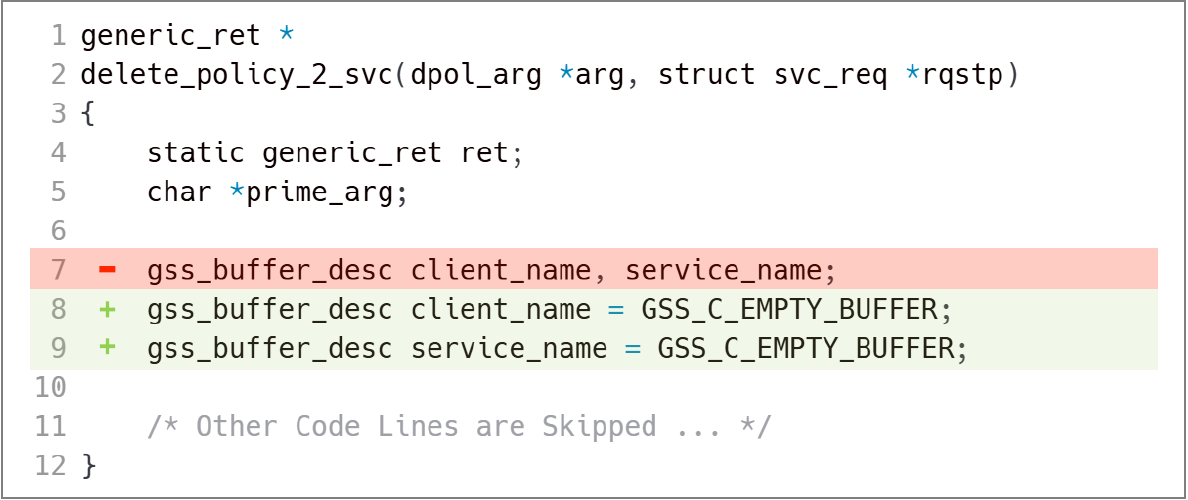}
        \caption{Example of a model-generated patch with \( EM = 1 \).}
        \label{fig2:external_test}
    \end{subfigure}
    \begin{subfigure}[b]{0.45\textwidth}
        \centering
        \includegraphics[width=\textwidth]{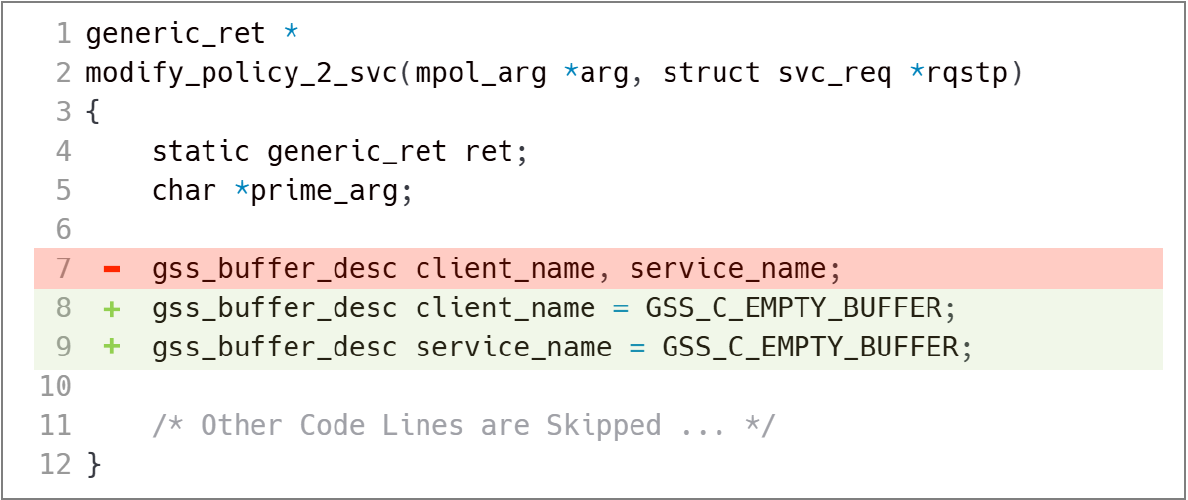}
        \caption{Example of overlapping data in the training set.}
        \label{fig2:external_train}
    \end{subfigure}
    \Description{Overlapping data caused by random splitting.}
    \caption{(CVE-2015-8631) Overlapping data caused by random splitting. \autoref{fig2:external_test} and \autoref{fig2:external_train} were repaired in the same commit, but were split into training and test sets.}
    \label{overlap}
\end{figure}

\subsection{Overlapping Data Problem (RQ2)}
\label{sec:3.3}
We first examine whether AVR models are tested on truly unseen data. 
In machine learning, we say that there is overlapping data when identical or highly similar samples appear in both the training and test sets, leading to inflated performance estimates.
This issue was partially acknowledged in VulMaster~\cite{vulmaster}, which found that 60\% of VulRepair's~\cite{vulrepair} dataset contained duplicate samples. 
While VulMaster attempted to mitigate this issue by removing duplicates, our analysis reveals that overlapping data still remains. 
We observe that patches from a single Git commit often share identical or near-identical repair patterns. 
Consequently, such patches become overlapping data when split and distributed across the training and test sets.
The inherent and unrealistic similarity between such patches gives the models an unfair advantage, leading to inflated performance estimates.

\autoref{overlap} demonstrates a typical scenario in which VulRepair and VulMaster produce a correct repair for a vulnerable function in the test set (\autoref{fig2:external_test}) with the help of overlapping data in the training set (\autoref{fig2:external_train}).
As shown in~\autoref{percentage of commit overlap}, this phenomenon is prevalent: over 85\% of repaired samples by VulRepair and VulMaster that achieved EM=1 are due to overlapping data from a single commit, suggesting that many successful patches may be partially attributed to memorizing training examples rather than learning genuine vulnerability patterns.
\begin{table}[t]
    \small
    \centering
    \caption{Commit overlap of correctly repaired samples generated by VulRepair and VulMaster}
    \label{percentage of commit overlap}
    \resizebox{\columnwidth}{!}{
\begin{tabular}{|c|c|c|c|}
    \hline
    \textbf{Approach} & \textbf{\# correct patches} & \textbf{\# commit overlap data} & \textbf{overlap rate (\%)} \\ \hline
    \hline
    \textbf{VulRepair~\cite{vulrepair}} & 295 & 253 & 85.76 \\
    \textbf{VulMaster~\cite{vulmaster}} & 330 & 291 & 88.18 \\ 
    \hline
\end{tabular}
}
\end{table}

\PP{Experimental Details} To evaluate how much this overlap inflates performance, we chronologically re-split the dataset to ensure the training data precedes the validation and testing data. 
This approach not only eliminates overlapping data but also aligns better with real-world usage, where newly reported vulnerabilities appear only after previous patches have been deployed. 
We kept the same sample counts for each subset to isolate the effect of data leakage from other confounding factors.
For this experiment, we only reevaluated VulRepair and VulMaster because VulAdvisor's data set does not accompany from which commit each patch sample is obtained.

\begin{table}[!t]
    \centering
    \footnotesize
    \renewcommand{\arraystretch}{1.2}
    \setlength{\tabcolsep}{5pt}
    \caption{Comparison of EM, BLEU, and CodeBLEU scores with and without overlap (denoted as Y and N, respectively) for VulRepair and VulMaster. The drop in evaluation scores is shown in parentheses.}  
    \label{tab:sorting}
    \begin{tabular}{|c|c|c|c|c|}
        \hline
        \textbf{Overlap}&   Y &   N &   Y &   N \\
        \hline
        \textbf{AVR Model} & \multicolumn{2}{c|}{\textbf{VulRepair}}
        & \multicolumn{2}{c|}{\textbf{VulMaster}} \\
        \hline
        \hline
        \textbf{EM} & 18.30 & 4.96 (-13.34) & 20.47 & 5.15 (-15.32) \\
        \textbf{BLEU} & 56.60 & 46.50 (-10.10) & 59.81 & 50.62 (-15.32) \\
        \textbf{CodeBLEU} & 55.57 & 47.36 (-8.21) & 58.25 & 50.84 (-7.41) \\
        \hline
\end{tabular}
\end{table}

\PP{Impact of Removing Overlap}
\autoref{tab:sorting} compares the EM, BLEU, and CodeBLEU scores
with and without overlapping data.
We observe a substantial drop in all metrics once the data is chronologically split to avoid overlapping data.
For instance, VulRepair's EM drops from 18.30\% to 4.96\% (-13.34 percentage points), and VulMaster's EM decreases from 20.47\% to 5.15\% (-15.32 percentage points).
Similar trends are observed for BLEU and CodeBLEU scores, with a drop of about 9 percentage points on average, indicating that much of the previously measured success was driven by highly similar patches appearing in both training and test sets.

For this experiment, we chose the best validation EM model for evaluating the test EM. Unlike in the overlapping setting, the best validation EM model occurs at Epoch 8 for VulRepair and Epoch 4 for VulMaster when we train them using the training and validation set without overlapping data.
After these points, the EM scores even declined as the training loss approached zero.
This suggests that in the presence of overlap, overfit models achieve the reported performance because they encounter near-duplicate samples in training, validation, and test sets.


\begin{findingbox}
To evaluate the true capability of AVR models, the evaluation should be done in a non-overlap setting. The presence of overlapping data in training and test sets leads to inflated match-based metric scores, masking the true generalization capabilities of fine-tuned AVR models. 
\end{findingbox}
\begin{figure}[t]
    \centering
    \begin{subfigure}[b]{0.45\textwidth}
        \centering
        \includegraphics[width=1\textwidth]{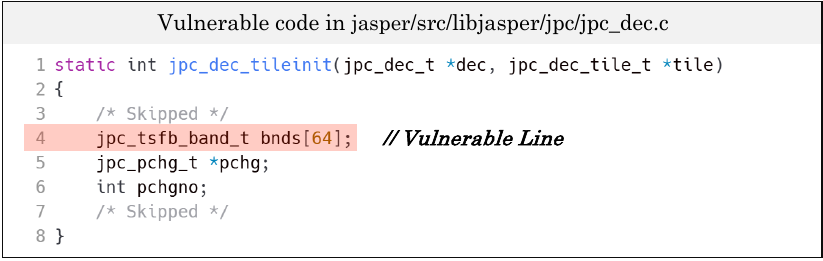}
        \subcaption{Example of a vulnerable code}
        \label{fig:vulnerable_code}
    \end{subfigure}
    
    \begin{subfigure}[b]{0.45\textwidth}
        \centering
        \includegraphics[width=1\textwidth]{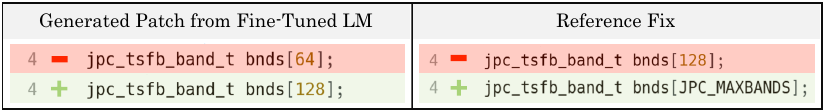}
        \subcaption{VulMaster-generated patch vs. reference fix}
        \label{fig:correct_patch}
    \end{subfigure}

     \begin{subfigure}[b]{0.45\textwidth}
        \centering
        \includegraphics[width=1\textwidth]{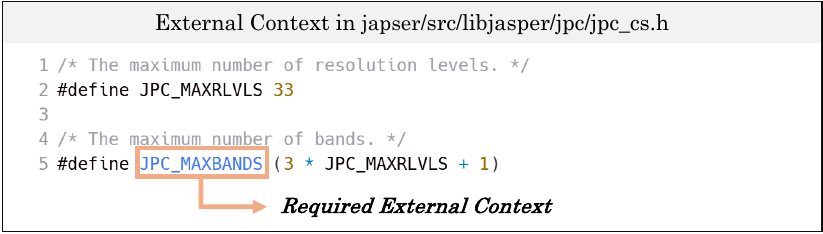}
        \subcaption{Example of external variable information necessary to generate the correct patch}
        \label{fig:external_context}
    \end{subfigure}
    \Description{Vulnerable code that requires an external variable to exactly match the reference fix.}
    \caption{(CVE-2016-9560) Vulnerable code that requires an external variable to exactly match the reference fix.}
    \label{fig:externalcontext}
\end{figure}

\section{Analysis on Evaluation Metrics for Fine-tuned AVR Models}
\PP{Overview}
In this section, we answer \textbf{RQ3}: How well do match-based metrics (i.e., EM, BLEU, CodeBLEU) correlate with key aspects of vulnerability repair, such as functional correctness and security?
Such match-based metrics have long been used in many natural and programming language processing tasks, including AVR, where the model-generated patches are compared with reference fixes (e.g., a developer's patch).
What remains unclear is whether these metrics are adequate for evaluating the capability of fine-tuned AVR models in composing functional patches that eliminate vulnerabilities.
For example, it is evident that a model-generated patch that exactly matches the reference fix (i.e., EM=1) is indeed correct.
However, the opposite is not necessarily true: it is premature to conclude that a patch with EM=0 is incorrect because a piece of code having different syntax or structure can still be semantically the same, and even the semantics of a correct patch may not be unique for a given vulnerability.
We also find that BLEU and CodeBLEU, although designed to mitigate the limitations of EM, still fail to reliably capture functional correctness and security properties.

We address \textbf{RQ3} by manually analyzing the dataset to assess the feasibility of models replicating reference fixes and measuring the correlation between (Code)BLEU scores and the correctness of the generated patches in two aspects: functionality and security.
What enables us to measure the correlation is the introduction of L-AVRBench~(\autoref{ss:bench-overview}), a test-based benchmark designed specifically for fine-tuned AVR models.
We also use L-AVRBench to comprehensively assess model-generated patches if they preserve functionality and eliminate vulnerabilities, and present the results in the next section~(\autoref{sec5}).

\subsection{Limitation of Match-Based Metrics (RQ3)}
\label{ss:rq3}

\begin{table}[t!]
    \small
    \centering
    \caption{The percentage of data samples that use external context (function calls or variables) to patch the vulnerable function for each C/C++ Vulnerability Dataset.}
    \label{tab:tb5}
    \resizebox{\columnwidth}{!}{
\begin{tabular}{|c|c|c|}
    \hline
    \textbf{Dataset} & \textbf{\# Samples} & \textbf{\# External Context Required} \\ 
    \hline
    \hline
    \textbf{BigVul+CVE-Fixes~\cite{bigvul,cvefixes}} & 5,800 & 4,076 (70.28\%)\\ 
    \textbf{MegaVul~\cite{megavul}} & 17,472 & 10,686 (61.16\%) \\ 
    \textbf{PrimeVul~\cite{primevul}} & 4,704 & 3,159 (67.15\%)  \\ 
    \textbf{VulAdvisor~\cite{zhang2024vuladvisor}} & 18,517 & 5,164 (27.88\%)  \\
    \hline
\end{tabular}
}
\end{table}

\PP{Exact Match}
The limitation of EM as a metric for evaluating AVR models comes from the existence of correct patches that differ in syntax or structure from the reference fix (i.e., EM=0).
Using three models, we generated a total of 7,560 patches by applying 36 attempts per vulnerability across 70 vulnerabilities in L-AVRBench. 
Among these, 131 patches were classified as correct according to the L-AVRBench criteria, and only four patches had EM = 1, as we elaborate in~\autoref{sec5}.

Further analysis shows that this issue is particularly critical when a reference fix relies on external variables or function calls that are inaccessible to the model.
\autoref{fig:externalcontext} illustrates this issue, where the reference fix (\autoref{fig:correct_patch}) introduces a variable (\texttt{JPC\_MAXBANDS}) defined in a different file (\autoref{fig:external_context}).
Unlike this, the AVR model substitutes the buffer size with a hard-coded value (128), which is a reasonable choice given its limited context.
Despite its validity, the EM metric classifies the model-generated patch as incorrect.
This shortcoming unfairly penalizes patches that achieve the same functionality through alternative implementations. 

The use of external contexts, such as preprocessor macros or function calls, is common in real-world C/C++ vulnerability datasets, as shown in \autoref{tab:tb5}.
For example, except for the VulAdvisor dataset, 60--70\% of cases require external context for a correct fix.
This underscores the inadequacy of evaluating vulnerability patches primarily through EM.

\begin{figure}[t]  
    \centering
    \begin{subfigure}[b]{\linewidth}
        \centering
        \includegraphics[width=\linewidth]{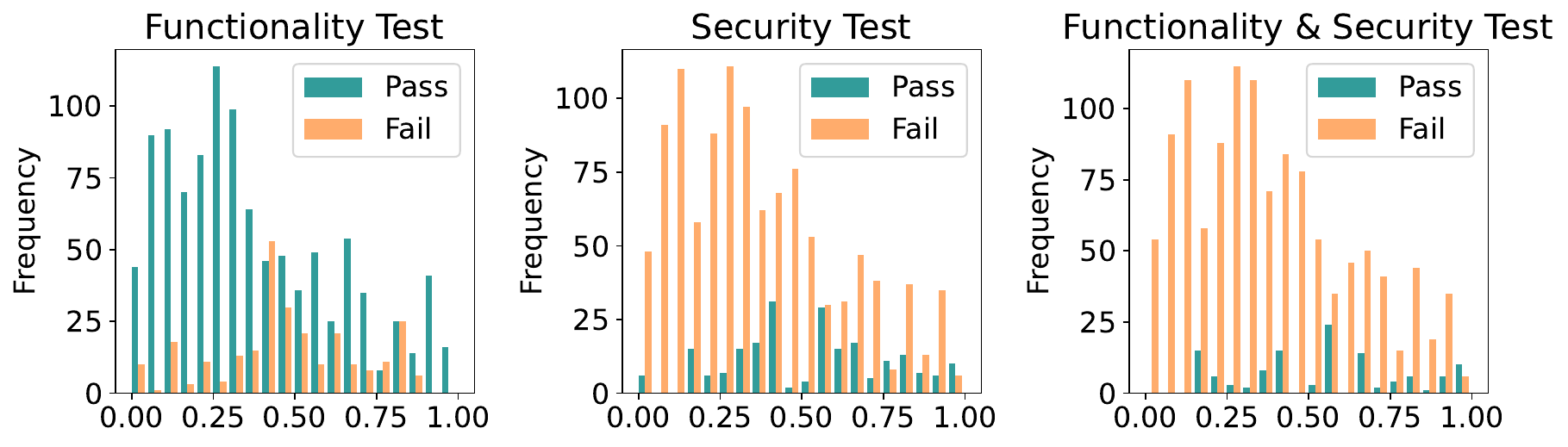}
        \caption{BLEU probability densities}
        \label{fig:bleu}
    \end{subfigure}
    \begin{subfigure}[b]{\linewidth} 
        \centering
        \includegraphics[width=\linewidth]{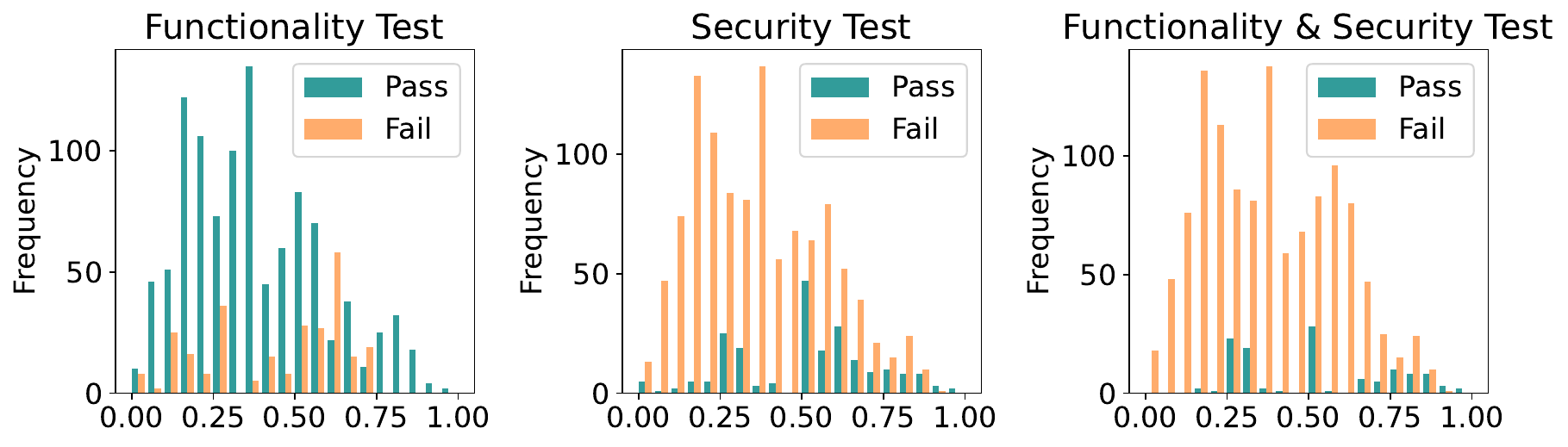}
        \caption{CodeBLEU probability densities}
        \label{fig:codebleu}
    \end{subfigure}
    \Description{The probability densities of existing metrics for passed (blue) and failed (yellow) patches under functionality and security tests of L-AVRBench are shown. The distribution overlap indicates that token-based similarity metrics fail to distinguish between correct and incorrect patches in terms of functionality or security.}
    \caption{The probability densities of existing metrics for passed (blue) and failed (yellow) patches under functionality and security tests of L-AVRBench are shown. The distribution overlap indicates that token-based similarity metrics fail to distinguish between correct and incorrect patches in terms of functionality or security.}
    \label{fig:distribution}
\end{figure}

\PP{BLEU and CodeBLEU}
Our quantitative analysis shows that BLEU and CodeBLEU also risk failing to capture functionality and security aspects of patch quality.
\autoref{fig:distribution} presents the probability density distributions of BLEU and CodeBLEU for patches that \emph{pass} or \emph{fail} the functionality and security tests in L-AVRBench.
We compute BLEU and CodeBLEU by comparing each generated patch with its reference solution, and observe substantial overlap between the score distributions for passing and failing patches. 
This indicates that higher BLEU or CodeBLEU scores do not reliably correspond to improved functional correctness or security.

\begin{findingbox}
EM can underestimate the capability of AVR models by failing to recognize semantically correct patches that differ textually from the reference fix. BLEU and CodeBLEU exhibit weak correlations with functional correctness and security, as their scores overlap significantly for both passing and failing patches. 
\end{findingbox}

The analysis presented in this section led us to advocate that the evaluation of fine-tuned AVR models should extend beyond EM, BLEU, and CodeBLEU.
The evaluation should incorporate test-based methods that explicitly verify functionality and security properties to mitigate the risk of underestimating or overestimating model performance, which we have observed in our analysis.

\subsection{L-AVRBench: A Test-Based Benchmark for Learning-Based AVR Models}
\label{ss:bench-overview}
This section introduces L-AVRBench, a test-based benchmark suite tailored for evaluating learning-based AVR models.
L-AVRBench evaluates patches using two types of tests: security tests and functionality tests.
The security tests use Proof-of-Concept (PoC) that triggers the vulnerabilities to determine if a patch nullifies the vulnerability or not, and use sanitizers (e.g., AddressSanitizer (ASan) \cite{asan}, UndefinedBehaviorSanitizer (UBSan)~\cite{polacek2014gcc}, MemorySanitizer (MSan) \cite{msan}, and ThreadSanitizer (TSan)~\cite{tsan}) or vulnerability-specific oracles to examine if the vulnerability is triggered or not by the PoC.
As functionality tests, L-AVRBench uses the regression tests or unit tests that each project accompanies.
By incorporating both security and functionality evaluations, L-AVRBench provides a direct assessment of AVR models, ensuring that patches nullify the vulnerability under the PoC while making the program pass all the unit tests.

\PP{Requirements}
L-AVRBench is designed to evaluate learning-based AVR models under realistic and reproducible conditions. To achieve this goal, we establish the following requirements for evaluating learning-based AVR models, which we capture as follows:
\begin{enumerate}[label=\textbf{R\arabic*}, leftmargin=2em, noitemsep, nolistsep]
    \item \textbf{Existence of a PoC Trigger.}  
    The benchmark must include a PoC input that deterministically triggers the vulnerability for the reproducibility of security tests.
    
    \item \textbf{Locality of Patches.}  
    The reference fix must be localized to a single function, aligning with the constraints of existing fine-tuned AVR models, which are typically trained to repair vulnerabilities at the function level.
    
    \item \textbf{Availability of Unit Tests.}  
    The project containing the vulnerable function must provide unit tests to assess that the patched code preserves the intended functionality, ensuring correctness beyond syntactic modifications. 
    Additionally, the reference fix must pass all provided functionality tests to be considered valid.
\end{enumerate}

L-AVRBench is constructed by combining vulnerabilities from two established sources, ExtractFix and Magma, as well as a set of real-world CVEs we manually curated.
These sources collectively ensure broad coverage across vulnerability types, realistic patching contexts, and alignment with our requirements.

\PP{Extractfix}
We begin by incorporating ExtractFix~\cite{gao2021beyond}, an existing benchmark for AVR evaluation, into L-AVRBench. ExtractFix provides real-world C/C++ vulnerabilities with developer-provided patches. However, not all samples meet our three requirements for test-based evaluation. Specifically, only 16 out of the original 30 ExtractFix samples include both unit tests and a working PoC, and have function-localized patches. We include only these 16 in L-AVRBench to ensure consistency with the benchmark design criteria.

\PP{Magma}
We adapt Magma~\cite{hazimeh2020magma}, a widely used benchmark to evaluate fuzzers~\cite{li2022pacmem, huang2024titan}, to enrich L-AVRBench.
Magma contains real-world security vulnerabilities that have been forward-ported to specific versions of open-source projects and utilizes canaries to verify that detected crashes originate from intentional vulnerabilities, ensuring reproducibility in fuzzer evaluations.
To make Magma suitable for AVR evaluation, we introduce several modifications that enhance both security and functionality testing.
First, we filter out projects whose unit tests are inconsistent.
PHP is the project that L-AVRBench does not adopt for this reason, because some of its test cases fail independently of the ported vulnerabilities.
Additionally, to ensure fairness in model evaluation, we remove conditional preprocessor directives that retain both the vulnerable and patched versions within the same file, keeping only the vulnerable version so that models generate patches solely based on the provided context.
Lastly, we refine test oracles by leveraging ASan and injected canaries to verify whether a model-generated patch successfully mitigates the vulnerability.
We also ensure that canaries are injected only after the model has generated the repair, preventing inadvertent hints about the correct fix during the repair process.
These modifications transform Magma into a robust benchmark for evaluating learning-based AVR, enabling comprehensive testing of both security and functional correctness in generated patches.

\PP{Hand-mined CVE Samples}
To further expand the benchmark and improve coverage of diverse vulnerability types, we manually curated additional real-world examples by mining publicly available vulnerability reports and patches from open-source C/C++ projects.
To focus on cases with clearly assigned CVE identifiers and ensure reproducibility, we limited our sources to (1) official CVE records~\cite{nvdsite} referencing commits in open-source repositories (e.g., GitHub), and (2) CVE-assigned bug reports from Google’s OSS-Fuzz project~\cite{ossfuzz}.
We reviewed each candidate manually to ensure that it satisfies the benchmark requirements (R1–R3).
Through this filtering and validation process, we obtained 27 samples, improving the L-AVRBench's coverage in terms of CWE IDs with real-world vulnerabilities.

\autoref{tab:tb6} gives an overview of L-AVRBench, detailing its diverse set of real-world vulnerabilities.
L-AVRBench consists of 70 vulnerabilities, each uniquely identified by a vulnerability ID, CVE ID, and CWE ID.
The dataset spans 15 open-source projects, covering a wide range of C/C++ software libraries, and contains a diverse set of vulnerabilities across 27 CWEs, including logic, concurrency, input validation, injection, resource management, and access control.

\begin{table}[!t]
    \centering
    \tiny
    \renewcommand{\arraystretch}{1.2}
    \caption{L-AVRBench Overview}
    \label{tab:tb6}
    \newcolumntype{M}[1]{>{\centering\arraybackslash}m{#1}} 
    \begin{tabular}{|C{0.85cm}|C{0.4cm}|C{1cm}|C{0.45cm}|C{0.05cm}|C{0.85cm}|C{0.4cm}|C{0.9cm}|C{0.45cm}|}
        \cline{1-4} \cline{6-9}
        \textbf{Program} & \textbf{ID} & \textbf{CVE} & \textbf{CWE} & & \textbf{Program} & \textbf{ID} & \textbf{CVE} & \textbf{CWE} \\
        \cline{1-4} \cline{6-9}
        
        \multirow{1}{*}{\centering Libpng} & LB1 & 2018-13785 & 369 190 & & \multirow{1}{*}{\centering Libxslt} & LB36 & 2019-18197 & 416 908 \\
        \cline{1-4} \cline{6-9}
        
        \multirow{19}{*}{\centering Libtiff} & LB2 & 2016-5314 & 787 & & \multirow{4}{*}{\centering OpenSSL} & LB37 & 2016-6309 & 416 \\
        & LB3 & 2016-10269 & 125 & && LB38 & 2017-3735 & 119 \\
        & LB4 & 2016-10269 & 125 & && LB39 & 2016-6302 & 20 \\
        & LB5 & 2015-8784 & 787 & && LB40 & 2016-7054 & 284 \\
        \cline{6-9}
        
        & LB6 & 2019-7663 & - & & \multirow{3}{*}{\centering SQLite} & LB41 & 2019-20218 & 755 \\
        & LB7 & 2017-11613 & 20 & && LB42 & 2013-7443 & 119 \\
        & LB8 & 2016-9273 & 125 & && LB43 & 2019-19926 & 476 \\
        \cline{6-9}
        
        & LB9 & 2016-5321 & 119 & & \multirow{6}{*}{\centering Libsndfile} & LB44 & 2011-2696 & 119 \\
        & LB10 & 2014-8128 & 787 & && LB45 & 2017-6892 & 119 \\
        & LB11 & 2014-8128 & 787 & && LB46 & commit a8ab5b3 & - \\
        & LB12 & 2016-10094 & 189 & && LB47 & commit a8ab5b3 & - \\
        \cline{6-9}
        
        & LB13 & 2017-7601 & 20 & & \multirow{4}{*}{\centering Binutils} & LB48 & 2018-10372 & 125 \\
        & LB14 & 2016-3623 & 369 & && LB49 & 2017-15025 & 369 \\
        & LB15 & 2017-7595 & 369 & && LB50 & 2017-9041 & 125 \\
        & LB16 & 2015-8668 & 787 & && LB51 & 2020-16590 & 415 \\
        \cline{6-9}
        
        & LB17 & 2016-10092 & 119 & & \multirow{2}{*}{\parbox{0.85cm}{\centering Libjpeg-\\turbo}} & LB52 & 2018-19664 & 125 \\
        \cline{1-4}
        
        \multirow{12}{*}{\centering Libxml2} & LB18 & 2017-9047 & 119 & && LB53 & 2012-2806 & 787 \\
        \cline{6-9}
        
        & LB19 & 2017-7375 & 611 & & \multirow{1}{*}{\centering Ffmpeg} & LB54 & 2024-36615 & 362 \\
        \cline{6-9}
        
        & LB20 & 2016-1762 & 119 & & \multirow{7}{*}{\parbox{0.85cm}{\centering Graphics\-magick}} & LB55 & 2017-8350 & 772 \\
        & LB21 & 2016-1839 & 125 & && LB56 & 2017-18271 & 835 \\
        & LB22 & 2016-1838 & 125 & && LB57 & 2018-5685 & 835 \\
        & LB23 & 2012-5134 & 119 & && LB58 & 2018-20189 & 20 \\
        & LB24 & 2017-5969 & 476 & && LB59 & 2019-11010 & 401 \\
        & LB25 & 2017-0663 & 787 & && LB60 & 2019-12921 & 77 \\
        & LB26 & 2025-24928 & 121 & && LB61 & 2025-27795 & 770 \\
        \cline{6-9}
        
        & LB27 & 2025-32415 & 125 1284 & & \multirow{9}{*}{\parbox{0.85cm}{\centering Image\-magick}} & LB62 & 2017-9501 & 617 \\
        \cline{1-4}
        
        \multirow{9}{*}{\centering Poppler} & LB28 & 2019-12293 & 125 & && LB63 & 2017-12140 & 400 681 \\
        & LB29 & Bug \#106061 & - & && LB64 & 2017-12643 & 770 \\
        & LB30 & Bug \#101366 & - & && LB65 & 2017-12665 & 772 \\
        & LB31 & 2019-7310 & 681 125 & && LB66 & 2017-14342 & 400 \\
        & LB32 & 2018-13988 & 125 & && LB67 & 2017-14533 & 772 \\
        & LB33 & 2018-10768 & 476 & && LB68 & 2017-18272 & 416 \\
        \cline{6-9}
        
        & LB34 & 2017-14617 & 20 & & \multirow{2}{*}{\centering jq} & LB69 & 2023-50246 & 120 \\
        \cline{1-4}
        
        \multirow{1}{*}{\centering Curl} & LB35 & 2018-1000301 & 126 & && LB70 & 2023-50268 & 120 \\
        \cline{1-4} \cline{6-9}
    \end{tabular}
\end{table}

\section{Evaluation with L-AVRBench (RQ4)} \label{sec5}

\begin{table*}[t]
    \centering
    \small
    \renewcommand{\arraystretch}{1.2}
    \caption{Results for EM, Compilation, Functionality, Security, and Functionality \& Security (Func.\&Sec.) evaluations across AVR models. \# Samp: Number of passed samples, \# ID: Number of passed IDs. (Dashes(-) represent 0 for readability.)}

    \label{tab:tab7}
     \resizebox{\textwidth}{!}{
\begin{tabular}{|C{1cm}|ccc|ccc|ccc|ccc|ccc||C{1cm}|ccc|ccc|ccc|ccc|ccc|}
        \hline
        \multirow{2}{*}{\makecell{ID}} & \multicolumn{3}{c|}{EM} & \multicolumn{3}{c|}{Compilation} & \multicolumn{3}{c|}{Functionality} & \multicolumn{3}{c|}{Security} & \multicolumn{3}{c||}{Func. \& Sec.}  
        &\multirow{2}{*}{\makecell{ID}} & \multicolumn{3}{c|}{EM} & \multicolumn{3}{c|}{Compilation} & \multicolumn{3}{c|}{Functionality} & \multicolumn{3}{c|}{Security} & \multicolumn{3}{c|}{Func. \& Sec.}\\
        
        & VR & VM & VA & VR & VM & VA & VR & VM & VA & VR & VM & VA & VR & VM & VA & &VR & VM & VA & VR & VM & VA & VR & VM & VA & VR & VM & VA & VR & VM & VA \\
        \hline\hline
        LB1 & - & - & - & - & - & - & - & - & - & - & - & - & - & - & - & LB37 & - & - & - & - & - & 10 & - & - & - & - & - & 3 & - & - & - \\
        LB2 & - & - & - & - & 2 & 19 & - & 2 & 19 & - & 2 & 17 & - & 2 & 17 & LB38 & - & - & - & - & - & 6 & - & - & - & - & - & - & - & - & - \\
        LB3 & - & - & - & - & - & 1 & - & - & 1 & - & - & - & - & - & - & LB39 & - & - & - & - & 14 & 14 & - & 14 & 14 & - & 2 & - & - & 2 & - \\
        LB4 & - & - & - & 1 & 26 & 4 & 1 & 26 & 4 & - & 23 & 1 & - & 23 & 1 & LB40 & - & - & - & - & 23 & 23 & - & 23 & 23 & - & 10 & - & - & 10 & - \\
        LB5 & - & - & - & - & - & - & - & - & - & - & - & - & - & - & - & LB41 & - & - & - & - & - & 3 & - & - & - & - & - & - & - & - & - \\
        LB6 & - & - & - & 1 & - & 22 & 1 & - & 22 & - & - & - & - & - & - & LB42 & - & - & - & - & - & - & - & - & - & - & - & - & - & - & - \\
        LB7 & - & - & - & - & - & 28 & - & - & 26 & - & - & 4 & - & - & 2 & LB43 & - & - & - & - & - & - & - & - & - & - & - & - & - & - & - \\
        LB8 & - & - & - & 27 & 27 & 28 & 2 & 5 & 28 & 25 & 22 & - & - & - & - & LB44 & - & - & - & - & 24 & 5 & - & 24 & 5 & - & - & - & - & - & - \\
        LB9 & - & - & - & - & 1 & 23 & - & 1 & 23 & - & 1 & - & - & 1 & - & LB45 & - & - & - & 30 & 15 & 27 & 12 & 7 & 27 & 18 & 5 & - & 2 & - & - \\
        LB10 & - & - & - & - & - & 26 & - & - & 25 & - & - & - & - & - & - & LB46 & - & - & - & - & - & 35 & - & - & 35 & - & - & - & - & - & - \\
        LB11 & - & - & - & - & - & - & - & - & - & - & - & - & - & - & - & LB47 & - & - & - & - & - & 16 & - & - & 15 & - & - & - & - & - & - \\
        LB12 & - & - & - & - & 35 & 3 & - & 35 & 3 & - & 23 & - & - & 23 & - & LB48 & - & - & - & - & - & 19 & - & - & 19 & - & - & - & - & - & - \\
        LB13 & - & - & - & - & - & 14 & - & - & 14 & - & - & 2 & - & - & 2 & LB49 & - & - & - & - & - & 11 & - & - & 11 & - & - & - & - & - & - \\
        LB14 & - & - & - & - & - & - & - & - & - & - & - & - & - & - & - & LB50 & - & - & - & - & - & 15 & - & - & 15 & - & - & 7 & - & - & - \\
        LB15 & - & - & - & - & 1 & - & - & 1 & - & - & - & - & - & - & - & LB51 & - & - & - & - & 4 & 17 & - & 4 & 17 & - & - & - & - & - & - \\
        LB16 & - & - & - & - & 5 & 19 & - & 5 & 19 & - & 4 & - & - & 4 & - & LB52 & - & - & - & - & - & 26 & - & - & - & - & - & - & - & - & - \\
        LB17 & 4 & - & - & 14 & 30 & 9 & 5 & 3 & 9 & 14 & 11 & - & 5 & 1 & - & LB53 & - & - & - & - & - & 32 & - & - & 23 & - & - & 9 & - & - & - \\
        LB18 & - & - & - & - & - & - & - & - & - & - & - & - & - & - & - & LB54 & - & - & - & - & - & 26 & - & - & 26 & - & - & - & - & - & - \\
        LB19 & - & - & - & - & 4 & 2 & - & - & 2 & - & 4 & - & - & - & - & LB55 & - & - & - & - & - & 3 & - & - & 3 & - & - & - & - & - & - \\
        LB20 & - & - & - & 30 & 22 & 26 & - & - & - & 3 & - & - & - & - & - & LB56 & - & - & - & - & - & - & - & - & - & - & - & - & - & - & - \\
        LB21 & - & - & - & 3 & 25 & 3 & - & 4 & - & - & 5 & - & - & 4 & - & LB57 & - & - & - & - & - & 15 & - & - & 5 & - & - & 6 & - & - & - \\
        LB22 & - & - & - & 18 & 12 & 2 & 16 & 1 & 2 & 2 & - & - & - & - & - & LB58 & - & - & - & - & 1 & 4 & - & 1 & 2 & - & - & - & - & - & - \\
        LB23 & - & - & - & - & - & - & - & - & - & - & - & - & - & - & - & LB59 & - & - & - & - & - & - & - & - & - & - & - & - & - & - & - \\
        LB24 & - & - & - & 8 & 7 & - & 8 & 7 & - & - & - & - & - & - & - & LB60 & - & - & - & - & 3 & - & - & 3 & - & - & - & - & - & - & - \\
        LB25 & - & - & - & - & 4 & - & - & 4 & - & - & 1 & - & - & 1 & - & LB61 & - & - & - & - & - & 15 & - & - & 15 & - & - & - & - & - & - \\
        LB26 & - & - & - & - & - & - & - & - & - & - & - & - & - & - & - & LB62 & - & - & - & - & 3 & 21 & - & 3 & 21 & - & - & - & - & - & - \\
        LB27 & - & - & - & - & - & 4 & - & - & 2 & - & - & - & - & - & - & LB63 & - & - & - & - & - & 5 & - & - & 2 & - & - & - & - & - & - \\
        LB28 & - & - & - & - & 18 & 3 & - & 18 & 3 & - & - & - & - & - & - & LB64 & - & - & - & - & - & 3 & - & - & 3 & - & - & - & - & - & - \\
        LB29 & - & - & - & - & - & 30 & - & - & 30 & - & - & - & - & - & - & LB65 & - & - & - & - & 9 & 29 & - & 9 & 29 & - & - & - & - & - & - \\
        LB30 & - & - & - & 19 & 9 & 33 & 19 & 9 & 33 & - & - & 2 & - & - & 2 & LB66 & - & - & - & - & - & 1 & - & - & 1 & - & - & - & - & - & - \\
        LB31 & - & - & - & - & - & 32 & - & - & 32 & - & - & - & - & - & - & LB67 & - & - & - & - & - & 1 & - & - & 1 & - & - & - & - & - & - \\
        LB32 & - & - & - & - & 9 & 21 & - & 2 & 21 & - & 8 & - & - & 1 & - & LB68 & - & - & - & - & - & - & - & - & - & - & - & - & - & - & - \\
        LB33 & - & - & - & 7 & 20 & - & 7 & 20 & - & - & 1 & - & - & 1 & - & LB69 & - & - & - & 9 & 23 & 9 & 9 & 23 & 9 & 6 & 11 & - & 6 & 11 & - \\
        LB34 & - & - & - & 29 & 20 & 18 & 29 & 20 & 18 & 6 & 3 & - & 6 & 3 & - & LB70 & - & - & - & - & 12 & 11 & - & 12 & 11 & - & - & - & - & - & - \\
        \cline{17-32}
        LB35 & - & - & - & 36 & 2 & 16 & 36 & 2 & 16 & - & 1 & - & - & 1 & - & \#Samp & 4 & - & - & 232 & 410 & 788 & 145 & 288 & 684 & 74 & 137 & 51 & 19 & 88 & 24 \\
        LB36 & - & - & - & - & - & - & - & - & - & - & - & - & - & - & - & \#ID & 1 & - & - & 14 & 31 & 52 & 12 & 29 & 46 & 7 & 18 & 9 & 4 & 15 & 5 \\
        \hline

\end{tabular}
}
\end{table*}

\subsection{Challenges and Our Approach}

We faced and overcame several challenges while evaluating the fine-tuned AVR models using L-AVRBench, which takes the test-based approach. The challenge primarily is in applying the generated patch to the original vulnerable functions for testing.

First, fine-tuned models are trained to generate patches with little to no surrounding context, making it difficult to apply them to the original vulnerable function for testing.
The lack of explicit context hinders seamless integration into the original code for functionality evaluation.
VulRepair and VulMaster generate only three tokens before and after the patch location, and having only three tokens does not uniquely specify the patch location in the function.
VulAdvisor produces the patch in the form of \texttt{diff} but does not provide the surrounding context, complicating the process of locating where each patch should be applied.
Second, VulRepair and VulMaster preprocess functions by tokenizing them and removing all newlines.
This preprocessing comes with several side effects, including the change of semantics for preprocessor directives that rely on line-based formatting, leading to compilation failures when the repaired code is evaluated using test suites.

To enable testing of these AVR models with a test-based benchmark, we retrained them to generate diffs that include both before and after contexts. The additional but required task of generating contexts may affect the model's performance negatively, as it increases the sequence length and complexity. However, our experiments show that the impact is not significant. Specifically, VulRepair, VulMaster, and VulAdvisor were reproduced with EM scores of 16.32, 18.6, and 20.30, with the additional tasks from 18.3, 20.47, and 23.3, respectively. We note that BLEU and CodeBLEU scores did not drop across all models. This may be attributed to the extended training duration, as we trained the three models for 10 to 15 additional epochs to ensure convergence. Additionally, we preserved the line-based formatting of vulnerable functions, ensuring compatibility with compilation, which is crucial for functionality and security testing. Lastly, we removed all the same CVEs introduced in L-AVRBench from the training set so that the model's evaluation results are not inflated.

\subsection{Experimental Results}
\PP{Main Results}~\autoref{tab:tab7} presents the evaluation results of three  AVR models: VulRepair (VR), VulMaster (VM), and VulAdvisor (VA) across multiple criteria—Exact Match (EM), Compilation Test, Functionality Test, and Security Test.
We also measured BLEU and CodeBLEU scores for the generated patches and reported the results earlier in~\autoref{fig:distribution}, and we discussed them in~\autoref{ss:rq3}. We exclude the numbers here due to space constraints.
The table summarizes the number of successful repairs per vulnerability ID, considering different evaluation aspects.
For each vulnerability ID, we generated 36 repair candidates per model, totaling 2,520 unique patches (\#Samp). 
Given that the quality of the generated patches can vary depending on the beam size, we experimented with multiple beam sizes: 1, 5, 10, and 20. 


Out of the 24 passed IDs, only one case exactly matched the developer-provided patch.
This suggests that while the models can generate correct patches, they often deviate in token-level structure from the original fixes, reinforcing the issue of only relying on EM as an evaluation metric.
The compilation success cases vary across models, with VulAdvisor (VA) achieving the highest overall compilation rate (788 successful cases), followed by VulMaster (VM) (410 cases) and VulRepair (VR) (232 cases). In terms of functionality, where a repair must preserve the intended program behavior, VA again performs best, passing 684 samples and 46 vulnerability IDs, while VM and VR pass 288 and 145 samples, covering 29 and 12 IDs, respectively. This suggests that VA is more effective at generating functionally correct repairs, potentially due to the size and diversity of its training dataset, which allows it to generalize better across different codes.

However, when considering security, which evaluates whether the generated patch successfully eliminates the underlying vulnerability, VM outperforms the other models, achieving 137 successful cases and passing 18 vulnerability IDs. In comparison, VA and VR pass 51 and 74 samples, covering 9 and 7 IDs, respectively. VM's superior performance in security-aware repair could be attributed to its ability to handle longer input sequences, incorporate AST representations, leverage CWE-specific information, and utilize repair knowledge from exact or related CWE-IDs.

\begin{findingbox}
Using L-AVRBench, we observe that VM resolves the most vulnerabilities (15 IDs), followed by VA (5 IDs), while VR resolves the fewest (4 IDs). The advantage of using a test-based evaluation is that it provides a more detailed comparison of models when they do not exactly match the reference fix.
\end{findingbox}

\begin{table}[t]
    \centering
    \footnotesize
    \caption{Authors' manual evaluations of AVR model patches.}
    \renewcommand{\arraystretch}{1.1}
    \label{tab:table_9}
\begin{tabular}{|c|l|l|l|c|l|l|}
\cline{1-3} \cline{5-7}
\textbf{Model}     & \textbf{ID}  & \textbf{Plausible} &  & \textbf{Model}     & \textbf{ID}  & \textbf{Plausible} \\ 
\cline{1-3} \cline{5-7} 

\multirow{4}{*}{VR} & LB17 & R   &  & \multirow{6}{*}{VM} & LB32 & NS \\
                    & LB34 & NR  &  &                      & LB33 & R \\
                    & LB45 & NR  &  &                      & LB34 & NF \\
                    & LB69 & R   &  &                      & LB35 & NF \\
\cline{1-3} 
\multirow{8}{*}{VM} & LB2  & NF  &  &                      & LB39 & NF \\
                    & LB4  & R   &  &                      & LB40 & NS/NF \\         
                    & LB9  & NF  &  &                      & LB69 & R \\
\cline{5-7}
                    & LB12 & NF  &  & \multirow{5}{*}{VA}  & LB2  & NF \\
                    & LB16 & NF  &  &                      & LB4  & NF \\
                    & LB17 & NF  &  &                      & LB7  & NR \\
                    & LB21 & R   &  &                      & LB13 & NR \\
                    & LB25 & NR  &  &                      & LB30 & NF \\
\cline{1-3} \cline{5-7}
\end{tabular}

\begin{tabular}{clllcll}
\multicolumn{3}{l}{{\underline{\textbf{R}}}: \underline{\textbf{R}}easonable} & &
\multicolumn{3}{l}{{\underline{\textbf{NF}}}: \underline{\textbf{N}}ot \underline{\textbf{F}}unctionally Correct} \\
\multicolumn{3}{l}{{\underline{\textbf{NS}}}: \underline{\textbf{N}}ot \underline{\textbf{S}}ecure} & &
\multicolumn{3}{l}{{\underline{\textbf{NR}}}: \underline{\textbf{N}}ot \underline{\textbf{R}}easonable} \\
\end{tabular}
\end{table}

\subsection{Manual Analysis}

\PP{Motivation}
A persistent challenge in AVR is that functionality tests often fail to ensure patch correctness. For example, Qi et al.~\cite{LeGoues2012GenProg} showed that despite GenProg claiming to fix 55 bugs~\cite{austin2021program}, only two were later verified as correct under deeper analysis. 
Similarly, prior work on learning-based AVR manually examined patches that had successfully passed functionality tests and found that nearly half were ultimately deemed not reasonable due to inadequate regression testing~\cite{pearce2023examining}. 
Moreover, proving the complete absence of vulnerabilities in generated patches remains difficult. Our evaluation relies on PoCs to validate repairs, which only confirm that a particular failure is addressed and do not guarantee that all instances of the vulnerability are eliminated.
There remains a possibility that the patch only provides a partial fix, leaving some aspects of the vulnerability unresolved.

\PP{Results}
We complement our automated evaluation with a manual analysis to gain deeper insights into the correctly classified patches by classifying the generated patches into four categories as follows. 
\textbf{Reasonable (R)} patches are either semantically equivalent to the reference fix or effectively address the vulnerability while differing in implementation from it, providing a valid and secure fix. 
\textbf{Not Functionally Correct (NF)} patches successfully remove the vulnerability but introduce unintended changes to the function’s expected behavior, potentially breaking functionality.
In contrast, \textbf{Not Secure (NS)} patches only partially resolve the vulnerability, leaving some attack vectors unaddressed. 
Lastly, \textbf{Not Reasonable (NR)} patches neither fully resolve the vulnerability nor preserve its intended functionality. 
This classification helps us understand the true capability of the fine-tuned AVR models.

To conduct this analysis, we manually reviewed all 131 patches that passed both functionality and security tests across all models.
Among these 131 patches, we report results for only one representative patch per ID in \autoref{tab:table_9}, specifically the patch with the best manual analysis outcome, since multiple patches from a single ID can pass both tests.
In one case, the patches generated for a single ID included both NF and NS variants, while other IDs typically fell into only one of these two categories. Since both functionality and security are critical, we conservatively reported this ID as NS/NF.

Among the 24 IDs that passed both functionality and security tests, 6 were classified as R, successfully eliminating the vulnerability while preserving original behavior. 12 were NF, introducing unintended functional changes despite eliminating the vulnerability. 2 were NS, likely blocking only the tested PoC input while leaving other exploit paths unaddressed. The remaining 5 were NR, neither fully resolving the vulnerability nor maintaining functional correctness.

Despite the presence of falsely classified samples, the trend in our manual evaluation confirms that VulMaster demonstrates the strongest overall repair capability, by producing 4 reasonable patches. Next, VR generated 2 reasonable patches, whereas VA failed to generate any, often producing patches that removed the vulnerability at the cost of functional correctness.

\begin{figure}[!t] 
    \centering
    \begin{subfigure}[b]{0.45\textwidth}
        \centering
        \includegraphics[width=\textwidth]{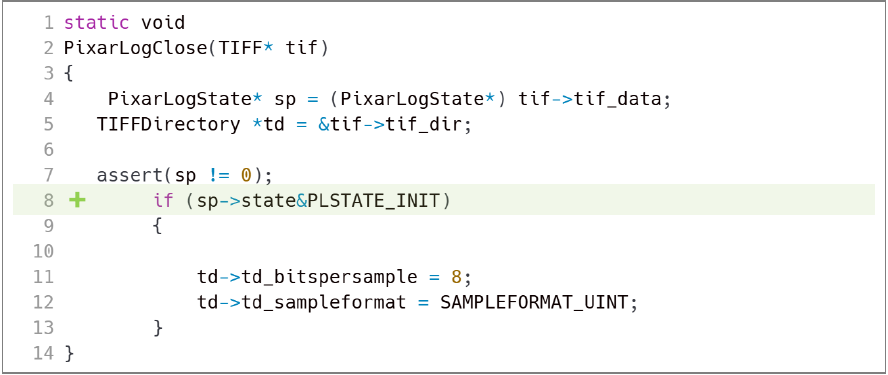}
        \caption{Reference fix.}
        \label{fig:cs_external_test}
    \end{subfigure}
    \begin{subfigure}[b]{0.45\textwidth}
        \centering
        \includegraphics[width=\textwidth]{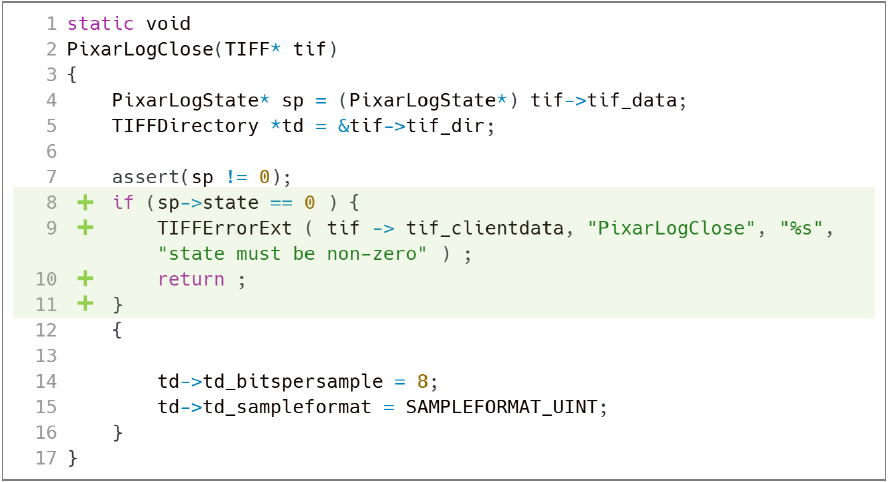}
        \caption{Repair patch generated by VulMaster.}
        \label{fig:cs_external_train}
    \end{subfigure}
    \Description{(LB4) Correctly repaired example when the reference fix requires an external variable.}
    \caption{(LB4) Correctly repaired example when the reference fix requires an external variable.}
    \label{case_study_1}
\end{figure}

\section{Discussion}
\subsection{Case Study}

In addition to the quantitative comparison using a test-based benchmark, L-AVRBench, we examine whether the test-based approach addresses one of the limitations of match-based metrics that we mentioned earlier: the unfair penalization of repairs that do not utilize external variables or function calls.
\autoref{case_study_1} shows such an example where the reference fix adds a conditional check using an external variable \texttt{PLSTATE\_INIT} that is defined as 1 in a different file.
Although this variable is not provided with the vulnerable code, VulMaster-generated repair applies a similar check \texttt{if (sp->state == 0)}, without explicitly referencing \texttt{PLSTATE\_INIT}. 
The patch is also classified as a reasonable repair by manual analysis because \texttt{sp->state} can only take values 0 or 1 in this function.
The example showcases that the test-based benchmark can evaluate the quality of generated patches even when they do not exactly match the reference fix.

\subsection{Guidelines and Implications for Future Research}
Based on the lessons learned from our study, we propose the following three guidelines for future research on evaluating AVR models.
First, avoid overfitting to superficial or non-vulnerability-related patterns that inflate match-based scores without reflecting true repair capability. Fine-tuned models must learn to generalize beyond training distributions to reliably fix unseen, real-world vulnerabilities.
Second, prevent data leakage through careful dataset splitting by ensuring that validation and test sets do not overlap with the training set. Apparent performance gains in such cases often reflect memorization of patches rather than true generalization. We recommend using a chronological order to avoid overlapping samples between the training and test sets, ensuring a fair and realistic evaluation.
Third, adopt test-based benchmarks such as L-AVRBench to assess whether generated patches are functional and secure. Test-based evaluation can also reveal trade-offs during training, where models may improve in one dimension (e.g., security) at the expense of another (e.g., functionality).
Together, we believe that these guidelines could lead to more robust, secure, and deployable automatic repair systems that go beyond memorization and toward truly generalizable vulnerability repair.

\subsection{Threats to Validity}
\PP{Internal Validity}
A common source of systematic bias in empirical studies on learning-based techniques is the choice of hyperparameters, which can be tuned to influence results toward a preferred outcome. To mitigate this risk, we used the hyperparameter values specified by the original authors and their implementations~\cite{vulmaster,zhang2024vuladvisor}. For VulAdvisor, not much information was written or released about the patch generation model, so we used the same setting as VulMaster to ensure consistency with VulRepair and VulMaster. 
Another threat lies in evaluating fine-tuned AVR models with potentially inadequate functionality tests and a limited number of crash inputs for security validation. To mitigate the risk of misinterpreting the models' capabilities, we conducted a thorough analysis of all 131 patches that passed both functionality and security tests, ensuring a more comprehensive assessment. Given the error-prone nature of manual analysis, two authors independently double-checked the patches to enhance reliability.

\PP{External Validity}
A key threat to the external validity is the limited number of fine-tuned AVR models evaluated. To mitigate this, we selected SOTA fine-tuned AVR models, VulMaster and VulAdvisor, along with the first fine-tuned AVR model, which serves as a baseline for recent work~\cite{kishiyama2024improving, liu2024t, mastropaolo2024training}.
Another limitation is the restricted number of test-based cases used to evaluate the models' repair capabilities. To enhance the generalizability of our findings, we extended the ExtractFix benchmark using Magma and Hand-mined CVEs. However, due to the manual effort required to adapt the benchmark for fine-tuned AVR models, we evaluated the models on only 70 vulnerable cases. Despite this, the overall takeaway from this study remains unchanged: the evaluation of AVR models should be rethought. 
Future work could further improve external validity by expanding the benchmark with more diverse projects and vulnerability types.
\section{Conclusion}
In this work, we assess the generalization and test-based performance of SOTA fine-tuned AVR models.
First, SOTA AVR models show limited generalization despite strong performance under conventional evaluation, with notable drops under semantic preserving transformations due to reliance on spurious features (RQ1).
Next, the commonly used random data-splitting approach unintentionally inflates performance due to overlapping data. Using chronological splitting instead yields a more realistic evaluation and reveals a drop in performance (RQ2).
Moreover, widely used match-based metrics (EM, BLEU, and CodeBLEU) do not accurately reflect the true capability of AVR models in terms of functional correctness and security.
Passing and failing patches exhibit substantial score overlap, indicating that higher BLEU or CodeBLEU scores do not reliably predict correctness or safety (RQ3).
Finally, test-based evaluation with L-AVRBench supplemented by manual analysis indicates that VulMaster consistently outperforms the other models. 
While VulAdvisor and VulRepair show comparable benchmark results, VulRepair produces more plausible patches in manual inspection. Notably, only one sample ID had a case that exactly matched the reference fix (RQ4).
Overall, while fine-tuned AVR models have made significant contributions to automatic vulnerability repair, our work suggests that a more comprehensive evaluation framework is needed to better understand their real-world effectiveness. 
We hope our findings contribute to ongoing research in this field and help refine evaluation methods to further improve AVR models in the future. 

\section*{Data Availability}

The implementation, benchmark samples, Docker images, and evaluation scripts
are publicly available at
\url{https://github.com/rimwoohan/L-AVRBench}.

\clearpage
\bibliographystyle{ACM-Reference-Format}
\bibliography{references}


\clearpage

\end{document}